\documentclass[a4paper,alpha-refs]{aejstyles}

\journal{aej}

\usepackage{graphicx}
\usepackage{siunitx}
\usepackage{layouts}

\usepackage[left]{lineno}
\newcolumntype{b}{X}
\newcolumntype{s}{>{\hsize=.4\hsize}L}

\title{A Pilot Study from the First Course-Based Undergraduate Research Experience for Online Degree-Seeking Astronomy Students}

\author[1,2,\authfn{1}]{Justin Hom}
\author[1]{Jennifer Patience}
\author[1]{Karen Knierman}
\author[1]{Molly N. Simon}
\author[1]{Ara Austin}

\affil[1]{Arizona State University}
\affil[2]{University of Arizona}

\authnote{\authfn{1}jrhom@asu.edu}

\papercat{Astronomy Education Practice}

\runningauthor{Hom et al. 2024}

\jvolume{00}
\jnumber{0}
\jyear{2017}

\begin{document}

\begin{frontmatter}
\maketitle
\begin{abstract}
Research-based active learning approaches are critical for the teaching and learning of undergraduate STEM majors. Course-based undergraduate research experiences (CUREs) are becoming more commonplace in traditional, in-person academic environments, but have only just started to be utilized in online education. Online education has been shown to create accessible pathways to knowledge for individuals from nontraditional student backgrounds, and increasing the diversity of STEM fields has been identified as a priority for future generations of scientists and engineers. We developed and instructed a rigorous, six-week curriculum on the topic of observational astronomy, dedicated to educating second year online astronomy students in practices and techniques for astronomical research. Throughout the course, the students learned about telescopes, the atmosphere, filter systems, adaptive optics systems, astronomical catalogs, and image viewing/processing tools. 
We developed a survey informed by previous research validated assessments aimed to evaluate course feedback, course impact, student self-efficacy, student science identity and community values, and student sense of belonging. The survey was administered at the conclusion of the course to all eleven students yielding eight total responses. Although preliminary, the results of our analysis indicate that students’ confidence in utilizing the tools and skills taught in the course was significant. Students also felt a great sense of belonging to the astronomy community and increased confidence in conducting astronomical research in the future.
\end{abstract}


\end{frontmatter}


\section{Introduction} \label{sec:introduction}
Student enrollment in online courses and degree programs have increased substantially in the past decade due to the benefits available to both students and instructors \citep{cooper2019factors,simon2022}. Online courses are more easily accessible to individuals who cannot regularly attend a college campus in-person, including military personnel, full-time workers, and international students. This accessibility has contributed significantly to the equity and inclusion of STEM education, with more students able to explore career pathways in STEM that were not available before \citep{perera2017,mead2020}. In providing online STEM courses for students, there has been a critical need for active learning materials that engage students in critical thinking, and skills development \citep{simon2022}.

Course-based undergraduate research experiences (CUREs) introduce students to the techniques and practices of performing scientific research in a structured and organized class centered on an active and scientifically meaningful research question \citep{auchincloss2014}. In a similar manner, Quasi-CUREs or QCUREs \citep{wooten2018} incorporate some but not all elements of CUREs, going beyond traditional classroom laboratory experiences. In a broad array of disciplines, CUREs and QCUREs have been found to increase, improve, or support: students' confidence in and competence in conducting science \citep{wang2015edu,brownell2015,gasper2013}, students' mastery of content knowledge \citep{koretsky2012,farr2012,purcell2016}, students' interest in science-related careers \citep{ward2014,shaner2016}, critical thinking skills \citep{gasper2013}, collaborative attitudes and practices, and student engagement in the field itself \citep{shaner2016,koretsky2012,bhattacharyya2009}. In the field of astronomy, CUREs and QCUREs have been piloted and developed in in-person curricula (e.g. \citealt{wooten2018}). 
\cite{wooten2018} found that QCUREs were able to improve astronomy students' science contribution self-efficacy while simultaneously increasing their appreciation for the importance of scientific analysis, community, and collaboration. CUREs prepare students for academic careers, but also have the potential to introduce skills that can be useful outside of academia. 

In this work, we present our experience developing and teaching a CURE intended for online, degree-seeking astronomy students at Arizona State University. The primary goal of this course is to prepare students for future research courses and programs. Science research careers are one of the predominant career pathways for astronomy degree students, with many seeking application to M.S./M.A. and PhD programs after receiving their bachelor's degree. Undergraduate students with some experience in science research are also more likely to pursue postgraduate degree programs (e.g. \citealt{eagan2013making,adedokun2012understanding}). 
These students are also able to provide evidence of efficiency in conducting research in postgraduate program applications. For maximal utility on graduate school applications, research programs are ideally started by the undergraduate student's third or fourth year in the degree program. 
Thus a central goal of this course was to inform students how to seek out research opportunities as undergraduates and provide them with the necessary skills and tools to conduct research. 

Students in this CURE manipulated observational astronomy data directly, performing their own measurements and calculations. Students worked in small groups during synchronous online class sessions and asynchronously through virtual meetings and group channels in a Slack workspace, allowing them to engage with the course material collaboratively regardless of location, time zone, or full-time employee commitments. At the conclusion of the course, we administered a survey to assess the impact of the course on multiple affective learning outcomes (e.g. self-efficacy, sense of belonging, and science identity).
As this was a pilot study, we do not have comparisons of students' experiences between the start and conclusion of the course. While we acknowledge the preliminary nature of our research study, our results are indicative of the potential impact of CUREs on the online astronomy major population.
\section{Astronomy Online Degree Program Overview and Demographics} \label{sec:onlinedegreeprogram}
Arizona State University's (ASU) online Bachelor of Science in Astronomical and Planetary Sciences (APS) is the first such degree in the United States. The degree provides broad training in the scientific foundations required to understand and to communicate the fundamentals of astronomy, space exploration, and ongoing advances in the field. The program of study includes groundwork classes and labs in mathematics and physical sciences, as well as topical courses focused on diverse fields within astronomy and planetary science. The assignments and projects the students complete are designed to provide exposure to the engineering and computational tools and techniques used to carry out scientific analysis in a range of fields. Students are prepared for a wide variety of careers including K-12 STEM teaching positions, science and technology journalism and writing careers, and technical careers involving statistical data analysis and/or computer programming. Compared to the in-person Bachelor of Science in Astrophysics, the APS degree currently does not include the 400-level sequence of astrophysics courses.
Consequently, students who are interested in attending graduate school in Astrophysics are advised to double-major in Physics and consider taking additional advanced coursework beyond the degree and to seek outside research experiences.  

Although the APS degree only opened in Fall 2020, the student enrollment is over 300 students as of Spring 2023, which is much larger than ASU's in-person enrollment in Astrophysics. The current enrollment has increased by over 100\% from the previous year. The demographics of online degree students are distinct from the in-person students which enables the program to reach a new and diverse population of learners. The average age of the APS degree online students is 30 years old and 56\% are male and 44\% female. Among the APS students, 25\% have a military affiliation with 26 students on active duty. Eleven percent are first generation college students. ASU also has corporate partnerships with companies such as Uber and Starbucks for tuition coverage, and 25\% of the APS students are from these corporate partnerships. Considering any category with more than five students, the ethnicity breakdown of the APS students is: 64\% White, 16\% Hispanic, 5\% two or more groups, 4\% African American, and 4\% Asian American. 
More than half of the APS students are transfer students, some from other majors within ASU and others from outside ASU (community colleges or other universities). One-quarter of the students are enrolling in college for the first time. Students returning to college to earn a second Bachelor’s degree comprise 17\% of the student population. A small subset, 5\% of students, are readmissions to ASU, indicating that they were away from ASU for more than 8 semesters. The demographics data was collected from student application information and provided by the university in an aggregated format, so it is not feasible to identify any correlations between categories. 

Separately from the current study, students from both online astronomy and in-person astrophysics programs were surveyed in Spring 2023 \citep{mead2023}. This survey had a completion rate relative to the total enrolled student populations of 25\% and 34\% for online and in-person students, respectively. None the less, these results (Table \ref{tab:inpersoncompare}) provide some sense of the similarities and differences between students in these two degree programs, highlighting that the online APS program has an increased proportion of students older than 30 years old, are first-generation college students, and are employed at some level.

\begin{table*}[bt!]
    \centering
    \caption{Comparison of in-person Astrophysics and online APS degree demographics collected in the Spring 2023 semester to be presented in \cite{mead2023}. The comparison demonstrates the broader diversity present in the online APS degree program compared to the in-person Astrophysics program, with higher percentages of first generation college students, students older than the age of 30, and employed students present in the online APS program.}
    \begin{tabularx}{\linewidth}{s | c |c | c | c} 
\toprule
{Demographic} & {In-Person Astrophysics [N]} & {In-Person Astrophysics [\%]} & {Online APS [N]} & {Online APS [\%]}\\
\midrule
\textbf{Age} &  &  & &  \\
\midrule
18-22 & 61 & 82.4 & 9 & 11.4 \\
23-29&9&12.2&20&25.3\\
30-44&4&5.4&31&39.2\\
45-54&0&0.0&14&17.7\\
55+&0&0.0&5&6.3\\
N&74&&79&\\	
\midrule
\textbf{Gender}	&		&		&		&		\\
\midrule
Man	&	38	&	51.4	&	44	&	55.0	\\
Woman	&	27	&	36.5	&	31	&	38.8	\\
Non-Binary	&	9	&	12.2	&	5	&	6.3	\\
N	&	74	&		&	80	&		\\
\midrule
\textbf{First Generation College Status} & & & & \\
\midrule
First Generation College (FG) total	&	21	&	28.4	&	42	&	53.2	\\
Non-FG total	&	53	&	71.6	&	37	&	46.8	\\
N	&	74	&		&	79	&		\\
\midrule
\textbf{Employment Status} & & & &\\
\midrule
Full time	&	9	&	12.2	&	43	&	54.4	\\
Not currently working	&	30	&	40.5	&	17	&	21.5	\\
Military	&	0	&	0.0	&	6	&	7.6	\\
Part time	&	35	&	47.3	&	13	&	16.5	\\
Any work (incl. part time and military) \%	&	44 &	59.5	&	62 &	78.5	\\ 
N	&	74	&		&	79	&		\\

\bottomrule
\end{tabularx}

    \label{tab:inpersoncompare}
\end{table*}

\section{Course Design and Structure}
\label{sec:coursedescription}
\subsection{Course Design in the Context of a CURE}
We designed our course as a CURE following a set of qualities defined in \cite{auchincloss2014}:

\begin{enumerate}
    \item Use of scientific practices: The course content actively encouraged students to ask questions, form hypotheses, utilize astronomical research tools, gather and analyze real-world data, develop interpretations, and communicate conclusions.
    \item Discovery: The research project is actively ongoing, and seeks to answer critical questions in theories of star formation and multiplicity (see \S \ref{sec:researchproject}). By working on the project, the students actively contribute to overall progress in answering the research questions involved.
    \item Broader relevance and scientific impact: The students' involvement in the course and analysis of the datasets will be acknowledged in upcoming scientific publications related to the research project.
    \item Collaboration: The students worked closely in small groups throughout the entire course, allowing them to develop collaborative and communication skills in the context of scientific research.
    \item Iteration: Students developed their own interpretations of the data analyzed and presented their group work to the rest of the class, similar to methods described in \cite{mclaughlin2016increasing}. Students were encouraged to challenge the methods and interpretations presented by their peers, and from this feedback, were able to reflect and/or revise their approaches in answering research questions.
\end{enumerate}

We describe in the next section how these aspects were integrated into the specific design of the course.


\subsection{Implementation of CURE Framework}
The course was offered as a 3-unit class at the 200-level for APS majors. To maximize the population of students able to take this introduction to research, the only pre-requisites were completion of pre-Calculus and the ability to install the publicly-available SAOImage DS9 software on their computers. The course was delivered in a fully online modality, with a combination of synchronous and asynchronous components. The synchronous component of the class involved zoom-based class meetings of 1.5 hours of planned instruction time 2--3 times per week (scheduled to accommodate student work schedules) with a team of two instructors, one teaching assistant, and a class of eleven students. As this was the first offering of this course, we opted to have a smaller class size compared to other core curriculum for the online APS major. The smaller class size allows for more focused attention and direct feedback to individual students in the course and also facilitates open discussion and presentations among project groups at a more manageable class volume. Providing and obtaining direct feedback among peers also allows the students to gain iterative experience in conducting science. Synchronous instruction time often extended longer than the planned 1.5 hours due to student questions and instructor responses to those questions. 
The class sessions were conducted in a workshop format with brief explanations and demonstrations of research tools, followed by hands-on activities with the tools for the students to complete in breakout rooms as the instructional team circulated amongst the groups to answer questions. 
The students also worked independently in an asynchronous mode and were also placed in groups to enable student-scheduled synchronous meetings with classmates. Each student group prepared a report on the research background and their results in shared documents and databases. Overall, this allows us to foster the collaborative experiences of the course, with students working closely together regardless of their geographic location in a similar manner to astronomical research collaborations. 

The initial offering of the CURE titled {\it SES 294: Research Experience in Astronomical Imaging} was Summer (2022) and focused on students gaining experience with AO imaging data from a companion search project. The project allows students to interface with a current research project with no clear results or answers, simulating an authentic research experience and encouraging students to make their own assertions and conclusions from the data provided \citep{quinn2011}. The Summer session at ASU is a compressed 6.5 week timeline intended to cover the same material as a full 15 week Fall or Spring semester course. The students developed valuable experience with imaging through a quantitative background understanding of the limits of telescope resolution and sensitivity and practical skills in displaying, processing, and measuring astronomical images. The lab exercises that formed the basis of the CURE involved activities such as working with astronomical databases accessible through \href{http://simbad.u-strasbg.fr/simbad/sim-fid}{SIMBAD/Vizier}, planning observations with online tools such as \href{http://catserver.ing.iac.es/staralt/}{Staralt}, inspecting data with tools such as SAOImage DS9, and performing measurements and making plots with spreadsheet software and Python notebooks. Introductory curriculum in Python consisted of Google Colaboratory Notebooks and the textbook \textit{Python for Astronomers: An Introduction to Scientific Coding} \citep{pasha2019}. The class culminated in the students virtually attending a night of observing for the class research project at Keck Observatory through Zoom, with Keck Observatory staff providing an overview of the facility and answering students' questions. 

The primary intention of the course was to provide students an introduction to conducting astronomical research in a low-pressure environment. As a result, grading was performed on a pass/fail basis. 
Student progress was monitored throughout the course regardless, and feedback was given on assignments and group presentations to ensure all students achieved a similar level of competence in skills related to conducting research.

\section{Research Project and its Relation to the Course}
\label{sec:researchproject}
The project aims at measuring the frequency and properties of stellar and substellar companions of nearby early-type stars with an adaptive optics (AO) imaging survey. The specific instrument used to conduct the research was NIRC2, an instrument on the Keck telescope operating at infrared wavelengths.
The research project measurements represented empirical tests of brown dwarf formation scenarios. Measurements of companion brown dwarf brightnesses are critical in assessing the robustness of planetary and stellar formation models. Newly identified brown dwarf companions are also benchmark systems for testing analytic planetary/stellar atmosphere models due to the known ages of the primary stars. The key research topics of the project -- brown dwarfs, AO imaging, companion search, observation planning, and reporting results -- were used to introduce or reinforce key skills for the students throughout the CURE. This particular research project was chosen because the instructors who implemented this CURE were already involved as co-investigators. Other CUREs \citep[e.g.,][]{brownell2015,ward2014} have successfully implemented similar approaches.

\begin{figure*}
    \centering
    \includegraphics[width=\textwidth]{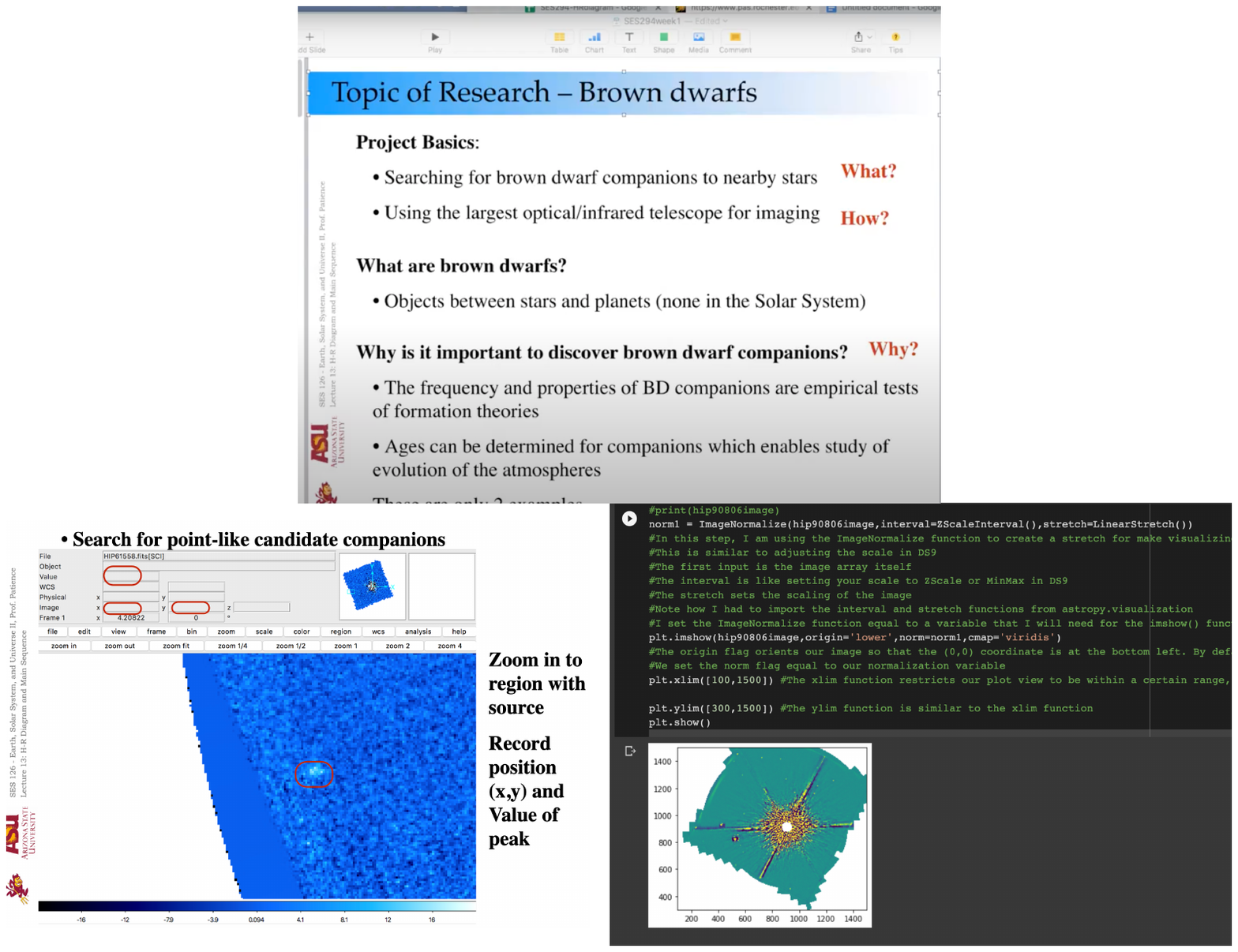}
    \caption{\textit{Top:} Screenshot from a slide of the lecture component of the course. This lesson presented an introduction of the research project and some background and motivation on the concept of brown dwarfs. \textit{Bottom left:} Screenshot of a window from SAOImage DS9 with a science image loaded. Red circles highlight important information for students to record, including the coordinate position of a candidate companion and its brightness value. \textit{Bottom right:} Screenshot from a Python notebook displayed in Google Colaboratory. The code block shown is a demonstration on importing science image files and displaying them using functions from the \texttt{matplotlib} \citep{matplotlib_v2.0.2} and \texttt{astropy} \citep{astropy2018} libraries.
    }\label{fig:SES294Visual}
\end{figure*}

Table \ref{tab:researchtopics} summarizes the links between the research topics of the project and how the research topics link to relevant content knowledge, research skills more broadly, and commonly used astronomical research tools more specifically.
The science topic of brown dwarfs is rarely covered in introductory astronomy courses. Therefore, covering the background material facilitated both a review of the basic properties of mass, radius, and temperature of stars, as well as an extension to brown dwarfs. The students explored the subject not through a textbook and homework assignments, but rather with guided searches for tables and formulae connecting physical and observable properties, followed by using web-based research tools to find, manipulate, and plot data. 
The utilization of adaptive optics in the project observations facilitated a comparison of the resolutions and sensitivity limits of ground-based and space-based instrumentation.
The students gained experience with querying observatory and scientific literature websites to identify and interpret information related to the scientific instruments utilized and the observations taken. The research images from the Keck telescope were the second epoch (second observation of the same targets) of a companion search, and this part of the project required a review of parallax and an introduction to proper motions and image display and processing fundamentals. The majority of the class involved experimental work related to this part of the research, as the students searched databases for information on the target proper motions and distances, calculated estimated motions of companion candidates between epochs, and performed measurements with SAOImage DS9 on the research images. In particular, the students measured count levels and identified positions of candidate companions. Figure \ref{fig:SES294Visual} presents screenshots of lecture material and interfacing with SAOImage DS9 and Google Colaboratory Python notebooks in order to view and analyze science images from the research project.

Since the research project was ongoing and had a virtual observing run at the Keck Observatory, the class had an observation planning component of the class in which the students used research databases and search tools such as SIMBAD to obtain coordinates, magnitudes, and proper motions of the target sample. With the date and time of the observing run, the students determined which targets were observable, how much pixel motion a background object would show relative to a target star, and what contrast level was needed to detect brown dwarfs with different spectral types. These calculations introduced students to coordinate systems and magnitude/filter systems while reinforcing background on moon phases magnitude differences in a real research program. Finally, the students worked in groups to prepare a research report with an overview of the background science, methodology, and results from the work performed throughout the class. This summary introduced the students to scientific writing and used the skills developed throughout the CURE to construct tables and figures with results. As part of the report, students prepared a Team Expertise section that provided an opportunity to reflect on and articulate the skills learned in the class and to describe their unique background experience (e.g. military service/training, current/prior career) as they would do for a telescope/grant proposal or a research internship application. 


\begin{table*}[bt!]
\caption{Research topics in the course with related aspects of relevant content knowledge, research skills, and astronomical tools utilized.}\label{tab:researchtopics}
\begin{tabularx}{\linewidth}{s | L | L | L} 
\toprule
{Research Topic} & {Relevant Content Knowledge} & {Research Skills} & {Astronomical Tools}\\
\midrule
\textbf{Brown Dwarfs} & Basic Properties - Mass/Radius & units, unit conversions & Google sheets  \\
 & Comparison with Stars & interpreting tables & research websites, Google sheets \\
 & H-R diagram & data plotting & Google sheets, Python \\
 & Basic Properties - Temperature/Wien's Law & calculations with formulae and conversions & Google sheets, Python \\
 & Brown dwarf research overview & literature searches, interpreting articles & NASA ADS, ASU library website, astrobites \\
 \midrule
 \textbf{AO Imaging} & Comparison of diffraction and seeing limit & calculations with formulae and conversions & Google sheets\\
 & Effects of the atmosphere on telescopes & literature searches, interpreting plots & NASA ADS, ASU library website\\
 & Overview of ground/space-based telescopes & interpreting technical manuals & observatory websites\\
 & Overview of adaptive optics & literature searches, interpreting plots & NASA ADS, ASU library website\\
 \midrule
 \textbf{Companion Search} & Image display/scaling & data plotting/manipulation & SAOImage DS9 \\
 & Basic image processing & literature searches, interpreting plots & SAOImage DS9, ASU Library website \\
 & Proper Motions & calculations with formulae and conversions & SIMBAD, Google sheets \\
 & Parallax & calculations with formulae and conversions & SIMBAD, Google sheets \\
 & Basic image measurements - position, counts & data plotting/manipulation & SAOImage DS9, Python \\
 \midrule
 \textbf{Observation Planning} & Coordinate systems - RA/Dec, Alt/Az & data plotting & SIMBAD, staralt \\
 & Moon phases & interpreting diagrams/plots & UNL simulation website, bright/dark/grey charts \\
 & Target positions & interpreting tables/databases & SIMBAD, Aladin viewer \\
 & Magnitude system & calculations with formulae & Google Sheets, Python \\
 & Filter systems & interpreting tables/plots & NASA ADS \\
 \midrule
 \textbf{Reporting Results} & Scientific writing & data presentation & Google docs, Google Sheets, Slack \\

\bottomrule
\end{tabularx}

\end{table*}

\section{Survey Questions/Methods}
\label{sec:surveyquestions}

After the conclusion of the course, we conducted a survey to assess the efficacy/impact of the class and gather feedback for incorporation into future iterations of the CURE. Specifically, we wanted to assess if we were able to achieve our desired short- and medium-term outcomes stated in \S \ref{sec:coursedescription} and obtain preliminary insight on longer-term outcomes such as enhanced science identity, career clarification, and persistence in science. We combined several research validated surveys in CUREs and chose questions with affective outcomes in-line with the course's goals (see \citealt{cooper2020,estrada2011,cooper2019factors,corwin2015}). Although some questions were replicated verbatim, several were modified to focus specifically on topics related to astronomy or the SES 294 CURE in general. Survey questions are categorized into nine categories: (1) general course feedback, (2) course impact, (3) self-efficacy/confidence in skills, (4) science identity, (5) science community values, (6) perception of research, (7) career aspirations, (8) sense of belonging, and (9) demographic information. General course feedback questions were dedicated to obtaining detailed comments about the structure of the course itself. In particular, we focused on students' impressions of the amount of course content and the course pacing to improve our pedagogical approaches for future iterations of the course. Course impact, self-efficacy, science identity, and sense of belonging questions were also selected to assess the impacts the course had on students.
The survey was approved by the Institutional Review Board of the Office of Research Integrity and Assurance (ORIA) at ASU, with conditions that included informing student survey participants that the survey and individual questions within the survey were optional, consent was obtained, and that individualized responses would not be released. 
The final survey was distributed through email and administered through Google Forms. Out of eleven enrolled students, eight students completed the survey. Responses were not required for any question.

\subsection{Survey Questions}
The full list of survey questions are listed in Tables \ref{tab:surveyquestions} and \ref{tab:surveyquestions2}.
Course feedback was collected in Questions 1-8, with Questions 1-5 containing Likert-style questions on a scale of 1-5 (except Question 5, which was on a scale of 1-4) and Questions 6-8 requesting long-form paragraph responses. 
Question 9 asked a Likert-style question on a scale of 1-5 and Question 10 asked a multiple choice question about student impressions on the overall impact of the course on their ability to conduct science, applicable skills, and career development. 
Students were also surveyed on their self-efficacy/confidence utilizing the tools and skills taught in the course (Questions 11-12). 
Several questions (13-16) were dedicated to understanding students' identity as an astronomer, and their interpretations of scientific community values. 
Questions pertaining to scientific community values were adopted from \cite{estrada2011}, and the question focused on the students' perception of scientific research was adapted from \cite{cooper2019ONLINE}. 
Related to understanding their identity as scientists, students were also queried on their career goals, selecting from multiple sectors (Questions 17-18).
Two questions specifically queried students on their sense of belonging within the astrophysics discipline, adapted from questions investigated in \cite{estrada2011}. 
Finally, the survey concluded with questions on a limited set of demographic information.
Due to the small size of the class, we did not request students' ethnicity, racial, disability, or veteran-status.

All Likert-style survey questions selected/adapted from previous research-validated surveys employ the same Likert scales as their original source \citep{estrada2011,corwin2015,cooper2019factors,cooper2020}. For Likert-style survey questions original to this work (Questions 1-4), the choice of a five point Likert scale was made to be roughly consistent with other survey questions selected from previous research-validated surveys. A five point Likert scale was also selected for Question 12, as Question 12 is an extension of Question 11 \citep{estrada2011} specifically focused on confidence in skills related to Python programming.

\begin{table*}[p]
\caption{Survey Questions with response means and standard deviations reported for every Likert-style question.}\label{tab:surveyquestions}
\begin{tabularx}{\linewidth}{L c c}
\toprule
{Question} & {Question Type} & {Mean$\pm$Std. Dev.}\\
\midrule
A. Course Feedback & & \\
\midrule
1. How would you describe your interest in topics/skills that you learned in the course?$^1$       & Likert-style & 5.0$\pm$0  \\
2. How would you describe the pace of the class overall?$^1$ & Likert-style & 3.0$\pm$0 \\
3. How would you describe the pace of the Python notebook workshops?$^1$ & Likert-style & 3.75$\pm$0.89 \\
4. How would you describe the amount of class material?$^1$  & Likert-style & 3.13$\pm$0.35 \\
5. How often did you do the following in SES 294?$^2$ & Likert-style & \\
\hspace{3mm}--discuss elements of my investigation with my classmates or instructors & Likert-style & 3.88$\pm$0.35 \\
\hspace{3mm}--reflect on what I was learning & Likert-style & 3.88$\pm$0.35 \\
\hspace{3mm}--contribute my ideas and suggestions during class discussions & Likert-style & 3.63$\pm$0.52 \\
\hspace{3mm}--provide constructive criticism to classmates and challenge each other’s interpretations & Likert-style & 3.50$\pm$0.76 \\
\hspace{3mm}--share the problems I encountered during my investigation and seek input on how to address them & Likert-style & 3.88$\pm$0.35 \\
6. Did you ever feel uncomfortable and/or unprepared at any time during the course? In what way, and why?$^1$ & Paragraph & N/A \\
7. What would you change about this course, if anything?$^1$ & Paragraph & N/A \\
8. Do you have anything else you wish to share about this course?$^1$ & Paragraph & N/A \\
\midrule
B. Course Impact & & \\
\midrule
9. How much do you think this SES 294 class impacted you in these aspects?$^1$ & Likert-style &  \\
\hspace{3mm}--Skills development & Likert-style & 4.63$\pm$0.52 \\
\hspace{3mm}--Ability to work in a research group & Likert-style & 5.0$\pm$0 \\
\hspace{3mm}--Confidence to approach faculty with questions & Likert-style & 5.0$\pm$0 \\
\hspace{3mm}--Motivation to work in a related field & Likert-style & 4.88$\pm$0.35 \\
10. What skills from this class do you see yourself using again in the future? (Check all that apply.)$^1$ & Multiple Choice & N/A \\

\midrule
C. Self-Efficacy/Confidence in Skills & & \\
\midrule
11. Please indicate how confident you are in your ability with the following:$^3$ & Likert-style &  \\
\hspace{3mm}--to use technical science skills (use of tools, instruments, and/or techniques)$^3$ & Likert-style & 4.63$\pm$0.52 \\
\hspace{3mm}--to figure out what data/observations to collect and how to collect them$^3$ & Likert-style & 4.63$\pm$0.52 \\
\hspace{3mm}--to use scientific literature and/or reports to guide research$^3$ & Likert-style & 4.63$\pm$0.52 \\
\hspace{3mm}--Google Sheets/Plotting$^1$ & Likert-style & 4.88$\pm$0.35 \\
\hspace{3mm}--Tools such as: Staralt/SIMBAD/ds9$^1$ & Likert-style & 4.75$\pm$0.46 \\
\hspace{3mm}--ASU Library \& reference searches$^1$ & Likert-style & 4.63$\pm$0.52 \\
\hspace{3mm}--Write a Project Report$^1$ & Likert-style & 3.75$\pm$0.89 \\
12. Please indicate how confident you are in your ability to do the following using the Python programming language:$^1$ & Likert-style & \\
\hspace{3mm}--Use Python Notebooks & Likert-style & 3.13$\pm$1.13 \\
\hspace{3mm}--assign variables a value & Likert-style & 3.50$\pm$1.14 \\
\hspace{3mm}--arithmetic with variables & Likert-style & 3.38$\pm$1.30 \\
\hspace{3mm}--create and manipulate arrays & Likert-style & 2.75$\pm$1.28 \\
\hspace{3mm}--create plots & Likert-style & 3.38$\pm$1.51 \\


\bottomrule
\end{tabularx}

\begin{tablenotes}
\item Likert-style questions were given on scales of 1-5 for Questions 1-4, 9, 11, and 12. Question 5 was given on a scale of 1-4. Each Likert-scale value corresponded to a phrase based on the context of the question. References for questions: 1. This work, 2. \citet{corwin2015}, 3. \citet{estrada2011}.
\end{tablenotes}
\end{table*}

\begin{table*}[p]
\caption{Continued: Survey Questions with response means and standard deviations reported for every Likert-style question.}\label{tab:surveyquestions2}
\begin{tabularx}{\linewidth}{L c c}
\toprule
{Question} & {Question Type} & {Mean$\pm$Std. Dev.}\\
\midrule
D. Science Identity, E. Science Community Values, F. Perception of Research, G. Career Aspirations, and H. Sense of Belonging & & \\
\midrule
13 (D). Please indicate the extent to which you agree with the statements below$^1$ & Likert-style & \\
\hspace{3mm}--I derive great personal satisfaction from working on a team that is doing important research & Likert-style & 6.0$\pm$0 \\
\hspace{3mm}--I have come to think of myself as an astronomer & Likert-style & 4.25$\pm$0.71 \\
\hspace{3mm}--The daily work of an astronomer is appealing to me & Likert-style & 6.0$\pm$0 \\
14 (E). Please rate how much the person in the description is like you.$^1$ & Likert-style & \\
\hspace{3mm}--A person who thinks it is valuable to conduct research that builds on the world’s scientific knowledge & Likert-style & 5.88$\pm$0.35 \\
\hspace{3mm}--A person who feels discovering something new in the sciences is thrilling & Likert-style & 5.88$\pm$0.35 \\
\hspace{3mm}--A person who thinks discussing new theories and ideas between scientists is important & Likert-style & 6.0$\pm$0 \\
15 (F). Scientific research is the type of research that is being done in faculty member research labs. Please indicate the extent you agree with the following statement: I conducted scientific research in the SES 294 course.$^2$ & Likert-style & 8.57$\pm$1.13 \\
16 (F). Briefly explain your answer to the above question about conducting scientific research.$^2$ & Paragraph & N/A\\
17 (G). To what extent do you intend to pursue a science-related research career?$^3$ & Likert-style & 4.50$\pm$0.76 \\
18 (G). Which sectors most closely match your career goals? (more than one answer is possible)$^4$ & Multiple Choice & N/A\\
19 (H). Please indicate the extent to which you agree with the statements below$^1$ & Likert-style & \\
\hspace{3mm}--I have a strong sense of belonging to the community of astronomers & Likert-style & 4.50$\pm$0.76 \\
\hspace{3mm}--I feel like I belong in the field of astronomy & Likert-style & 4.88$\pm$0.35 \\

\midrule
I. Demographics & & \\
\midrule
20. Are you currently working while completing your degree?$^4$ & Multiple Choice & N/A \\
21. If you are currently working, what sector do you work in? (multiple answers are possible)$^4$ & Multiple Choice & N/A \\
22. Age$^4$ & Multiple Choice & N/A \\
23. Gender$^4$ & Multiple Choice & N/A \\

\bottomrule
\end{tabularx}

\begin{tablenotes}
\item Likert-style questions were given on scales of 1-5 for 17 and 19. Question 14 was given on a scale of 1-6 and Question 15 was given on a scale of 1-10. Each Likert-scale value corresponded to a phrase based on the context of the question, except for Questions 15 and 17 which were purely on a numbered scale representing varying degrees of responses. References for questions: 1. \citet{estrada2011}, 2. \citet{cooper2019factors}, 3. \citet{cooper2020}, 4. This work.
\end{tablenotes}
\end{table*}

\section{Data Description} \label{sec:datadescription}
Survey results are anonymous and were collated into visual forms such as pie charts, bar charts, and tables. Pie charts were used if a question presented multiple descriptive options from which the students were only allowed to select one option. Most of these questions were Likert-style questions. Bar charts were used for questions where multiple options were allowed to be selected, or if the options for a question were purely a numerical ranking. Although Google Forms records individual responses, all results are presented collectively, with the exception of some quotations from paragraph responses to specific questions. 
 Questions 8, 17, and 18 had seven of eight responses, and Question 21 had only three responses. The lack of responses for Question 21 is due to the nature of the question, which asked for responses from students who were full-time employees only. While our sample size is small, we emphasize that this course was a pilot study, and our goal was to provide a motivation for developing similar courses for online degree programs and describing ideal and non-ideal approaches for such a curriculum design. The small size of the sample and course in general were also purposeful choices to increase the effectiveness of the course structure and to encourage greater individual interaction between instructors and students. 







\section{Survey Results} \label{sec:surveyresults}
A summary of survey results is presented in this section. Detailed aggregate survey results are described in Appendix \ref{appendix:fullsurveyresults}.

\subsection{General Course Feedback}


In response to Questions 1-4, all students ($N=8$) expressed high interest in the subject matter of the class and found the overall pace of course content was manageable. Seven of eight students found the amount of course material was manageable, while the remaining student stated the amount of course material was "somewhat little." For the Python coding section of the course, responses were mixed, with half of the respondents reporting a manageable pace and the other half reporting a fast or too fast pace. Responses to Questions 2 and 3 are summarized in Figure \ref{fig:Q2,3classpacev3}.

\begin{figure*}
\centering
\includegraphics[width=\textwidth]{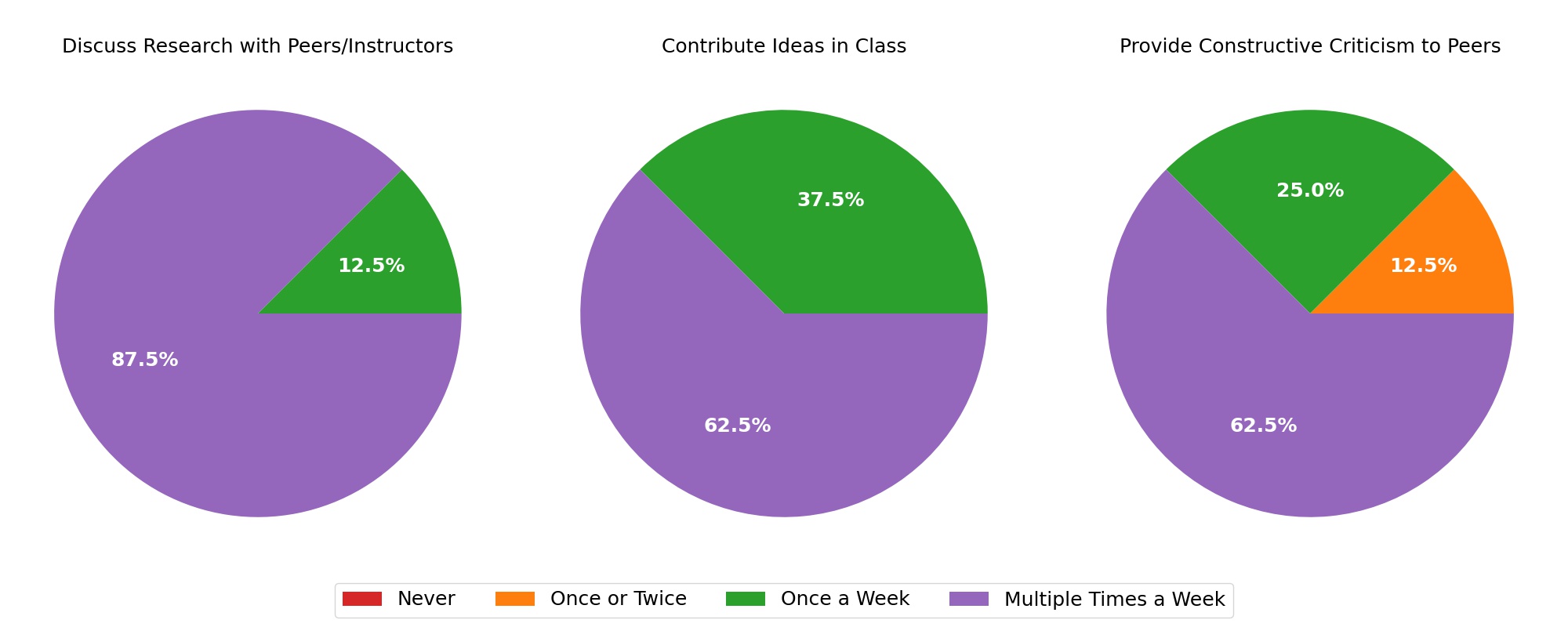}
\caption{A subset of results from Question 5: "How often did you do the following in SES 294?" The topics listed were "[discussing] elements of my investigation with my classmates or instructors," "[contributing] my ideas and suggestions during class discussions," and "[providing] constructive criticism to classmates and challenge each other’s interpretations." "[Reflecting] on what I was learning" and "[sharing] the problems I encountered during my investigation..." had identical results to "[discussing] elements of my investigation with my classmates or instructors." The frequency of which students performed these actions are encouraging, as these actions are commonly performed in STEM research and working environments.
}\label{fig:Q11:frequency}
\end{figure*}

Question 5, summarized in Figure \ref{fig:Q11:frequency}, asked students the frequency of which they performed certain actions within the course. Overall, 87.5\% of students were highly engaged within the course, with seven of eight respondents indicating they performed the following actions: interacting with their peers and instructors, reflecting on their own progress within the course, and seeking input on problems they encountered in their investigations multiple times a week. Five of eight respondents identified contributing to class discussions and providing constructive criticism to their peers multiple times a week, with the remaining only doing so once or twice a week. 

In general, students did not experience much discomfort in the course. One respondent to Question 6 stated that \textit{“after taking the course my confidence has gone up and I'm ready and eager to take on any research experience,”} while another stated, \textit{“The group work and teaching helped me through everything I struggled with.”} Two of eight respondents stated that the pace of the Python sections of the course was very fast, but one respondent stated that the Python curriculum would better prepare them for future classes.

A common theme among responses to Questions 7 and 8 regarding recommended changes to the course was the extension of the course to a longer length (8-15 weeks) as opposed to a Summer session (6 weeks). Three respondents stated a need for more time for the Python sections of the class, and that four class sessions in the last two weeks of the course was not sufficient. Additionally, two respondents expressed a desire for course content to be more evenly distributed throughout the CURE.

In the paragraph responses to Question 8, students expressed overall satisfaction and positive opinions about the course. No strictly negative responses were received. One respondent stated, \textit{“It gave me a real experience in what my future might hold and what its all about. My favorite part was the effort taken to make us feel there and involved even though it was an online course.”}


\begin{figure}[bt!]
\centering
\includegraphics[width=0.7\linewidth]{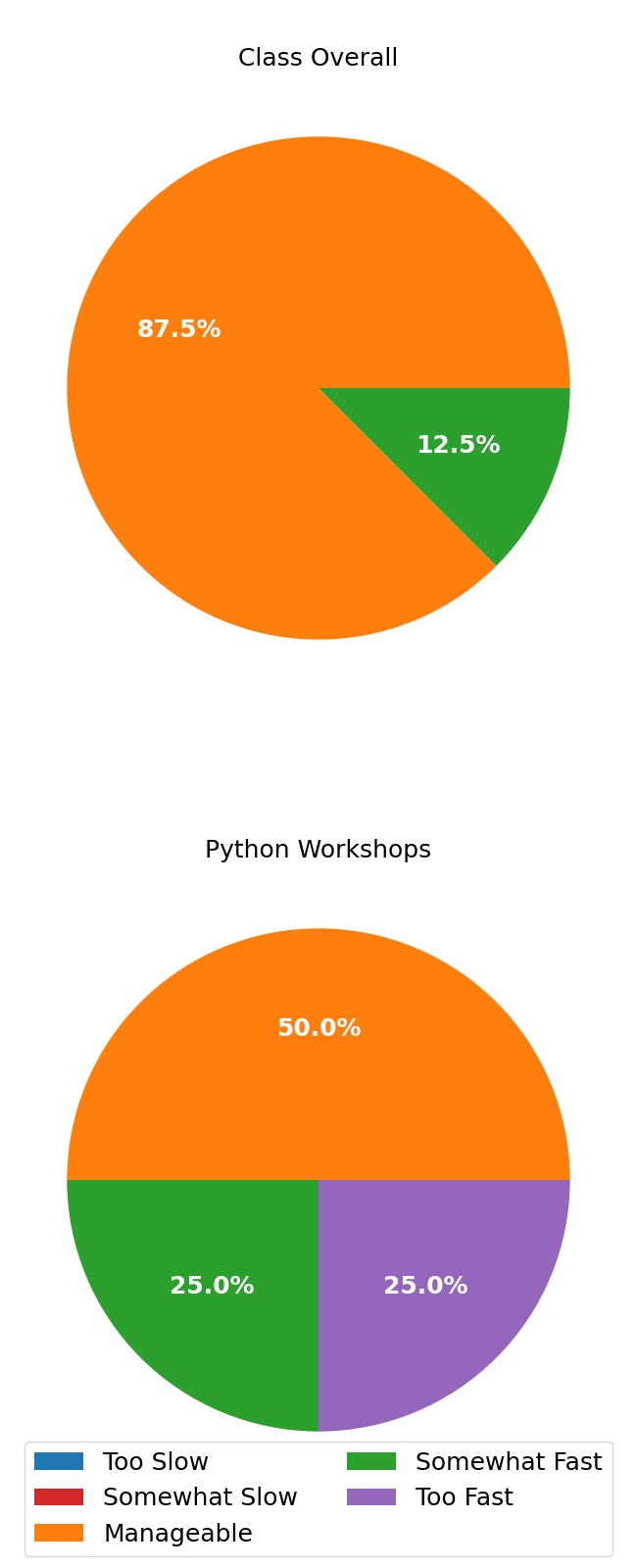}
\caption{\textit{Top:} Results from Question 2: "How would you describe the pace of the class overall?" \textit{Bottom:} Results from Question 3: "How would you describe the pace of the Python notebook workshops?" Overall, students found the pace of the course manageable but had more mixed responses regarding the pace of the Python workshops.
}\label{fig:Q2,3classpacev3}
\end{figure}


\subsection{Course Impact}

Question 9 asked students how much they thought the class impacted them across a variety of aspects. In the aspects of ability to work in a research group and confidence to approach faculty, all students rated the course's impact as "excellent." In motivating them to work in a related field, seven of eight rated the course's impact as "excellent" while the remaining student rated the impact as "good." Finally, five of eight students rated the course's impact on skills development as "excellent" while the remainder rated the impact as "good."

\begin{figure*}
\centering
\includegraphics[width=\textwidth]{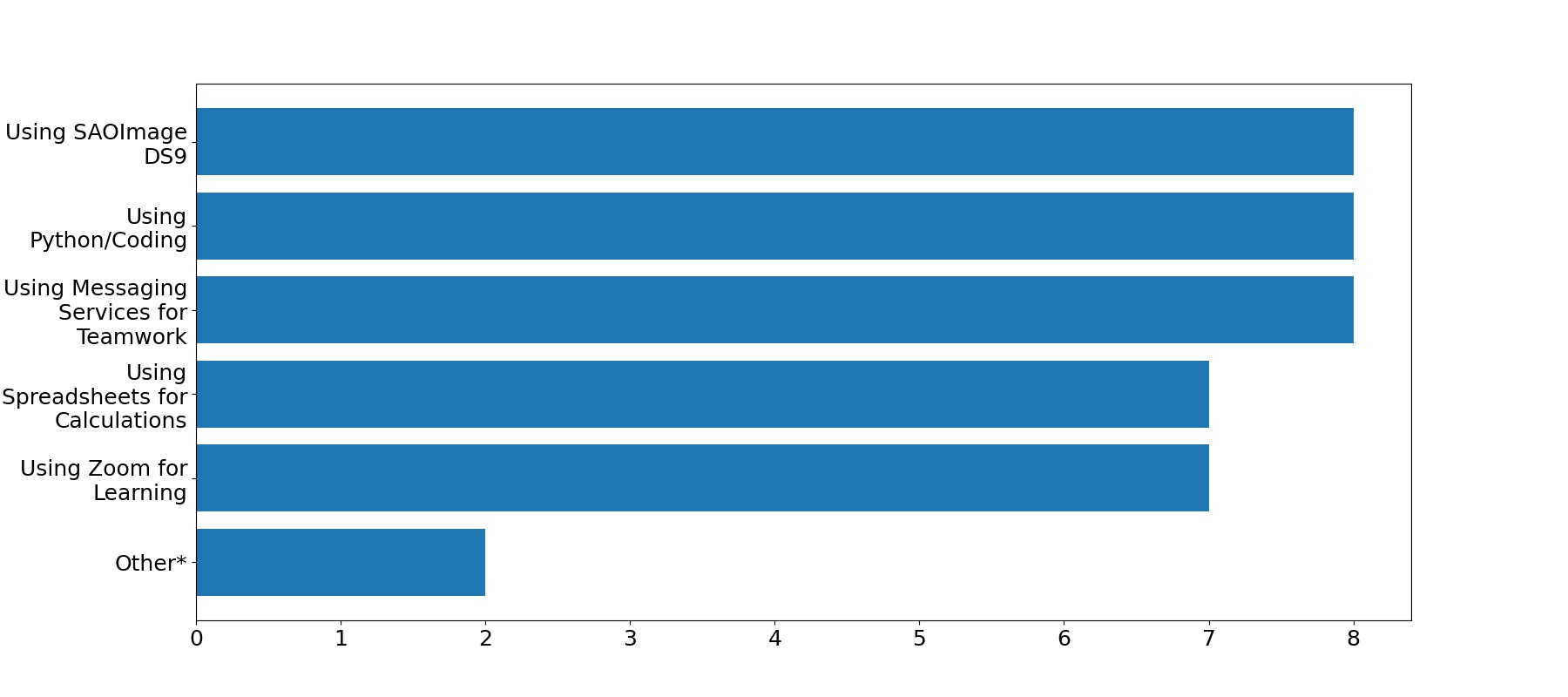}
\caption{Results from Question 10: "[Which] skills from this class do you see yourself using again in the future?" The possible options were "using SAOImage DS9 for examining astronomical images," "using Python or other coding," "using Slack or other messaging services to facilitate teamwork," "using spreadsheets for calculations (e.g., Excel or Google Sheets)," and "using Zoom for workshop style learning." *Two responses gave self-provided answers for skills that included "reading multiple scientific publications to look for relevant information to my research...Searching archived celestial objects observation data...[and] Planning of future observations" and "SIMBAD, staralt and NASA archives." These responses are categorized as "Other" in this Figure. Overall, students are of the opinion that the skills and tools they learned in this class will likely be helpful in the future.
}\label{fig:Q9:skillsfuture}
\end{figure*}

Question 10 asked students which skills they envisioned using in the future for courses, research opportunities, and careers. The question allowed for multiple responses, and the results of this question are shown in Figure \ref{fig:Q9:skillsfuture}. All students saw themselves using SAOImage DS9 for viewing astronomical images, using Python or another computer programming language, and using Slack or some other medium to facilitate teamwork in the future. 
Almost all students saw themselves using spreadsheets for calculations and using Zoom for workshop learning. 
Two respondents mentioned additional skills they would use in the future, including reading scientific journal articles to guide research, referencing archival data/observations, utilizing online tools/catalogues, and planning observation runs. 

\subsection{Self-Efficacy/Confidence in Skills}
\begin{figure*}
\centering
\includegraphics[width=\textwidth]{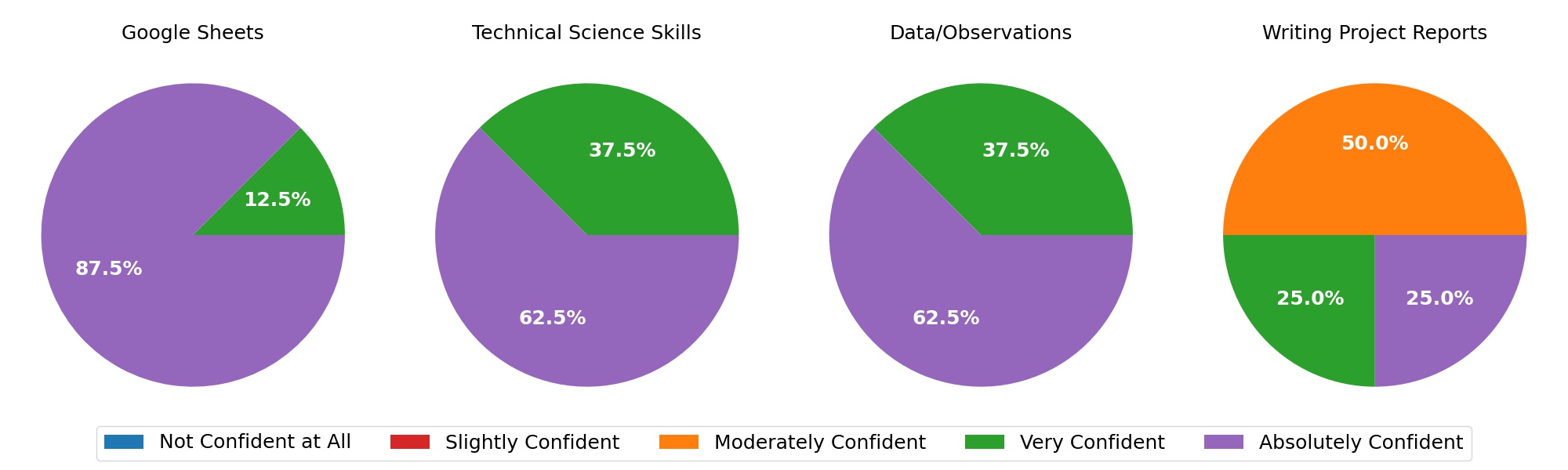}
\caption{A subset of results from Question 11: "Please indicate how confident you are in your ability with the following." The topics listed were "Google Sheets/Plotting," "to use technical science skills (use of tools, instruments, and/or techniques)," "[figuring] out what data/observations to collect and how to collect them," and "[writing] project reports." All topics not shown and described in Table \ref{tab:surveyquestions} have similar results to "Google Sheets/Plotting" and "Technical Science Skills." Student confidence in these skills is strongly encouraging, but more attention should be placed towards science writing and project reports in future iterations of the CURE.
}\label{fig:Q5:confidencev3}
\end{figure*}

Figure \ref{fig:Q5:confidencev3} summarizes student responses to Question 11, which queried students' self-efficacy/confidence in a variety of skills and tools taught throughout the CURE. Seven of eight students were "absolutely confident" in their ability to use Google Sheets. 
Six of eight students were also "absolutely confident" in using Online GUI tools, the ASU Online Library database, and understanding the nature of data/observations. 
Students were relatively more mixed on using scientific literature for guiding research and writing project reports, with four of eight students expressing only moderate confidence in writing reports specifically. 

\begin{figure*}
\centering
\includegraphics[width=\textwidth]{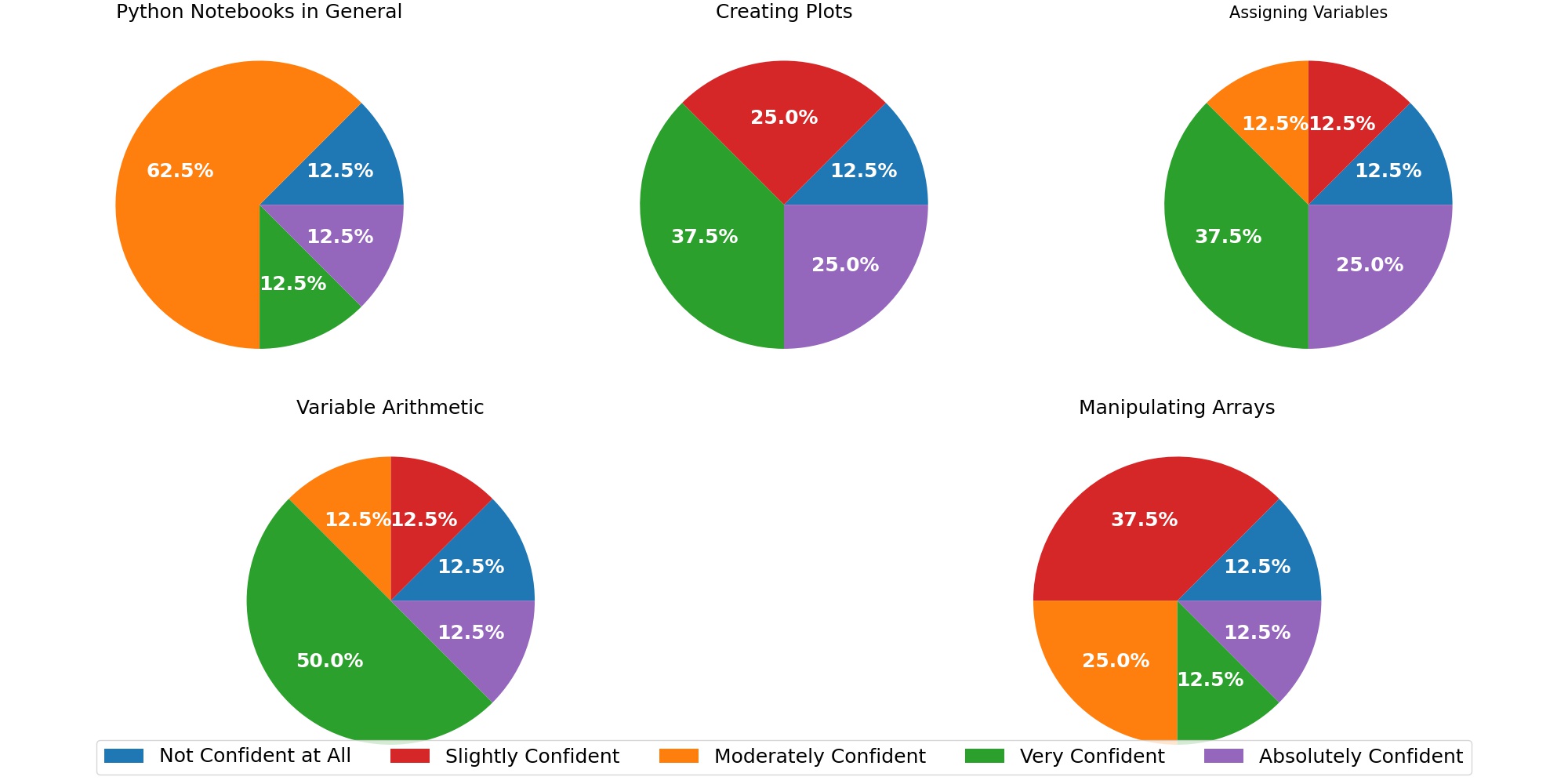}
\caption{Results from Question 12: "Please indicate how confident you are in your ability to do the following using the Python programming language." The topics listed were "[using] Python notebooks," "[creating] plots," "[assigning] variables a value," "[performing] arithmetic with variables," and "[creating] and [manipulating] arrays." Student confidence in Python skills was mixed, likely due to the nature in which Python lessons were organized in the CURE.
}\label{fig:Q6:Pythonconfidence}
\end{figure*}

The largest range in survey responses was given in Question 12 which specifically queried confidence in performing certain actions in Python. These results are summarized in Figure \ref{fig:Q6:Pythonconfidence}. Overall, students were less confident in performing actions in Python compared to other activities in the course. In particular, students were the most comfortable with assigning variables, performing arithmetic with variables, and creating graphs/plots. Students overall expressed less confidence in manipulating arrays and other large data structures. 


\subsection{Science Identity}

\begin{figure}[bt!]
\centering
\includegraphics[width=0.7\linewidth]{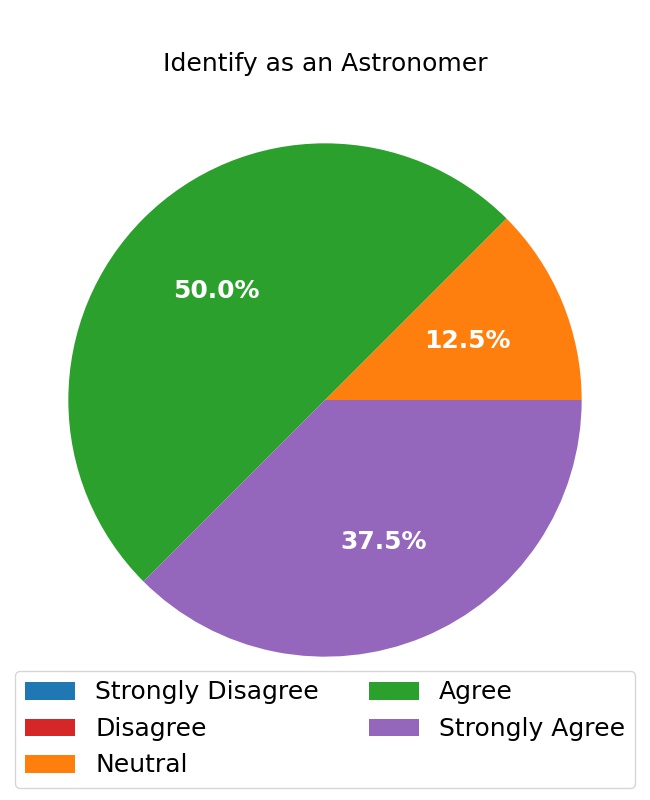}
\caption{A result from one of the sub-questions of Question 13: "Please indicate the extent to which you agree with the statements below: I have come to think of myself as an astronomer." All students "Strongly agreed" to the other sub-questions as stated in Table \ref{tab:surveyquestions2}. Overall agreement in a sense of identifying as an astronomer is encouraging for student retention within the field of astronomy.
}\label{fig:Q7:identity}
\end{figure}

Figure \ref{fig:Q7:identity} summarizes the results from Question 13, which asked students the extent at which they agree with statements related to science identity. All students strongly agreed that "the daily work of an astronomer is appealing" and that they "derive great personal satisfaction from working on a team that is doing important research." Seven of eight students agreed or strongly agreed that they identified as an astronomer.

\subsection{Science Community Values}

Question 14 queried students on their scientific community values. All students identified or strongly identified with scientific community values, which is also encouraging for students planning on pursuing research in the future.

\subsection{Perception of Research}

\begin{figure}[bt!] 
\centering
\includegraphics[width=\linewidth]{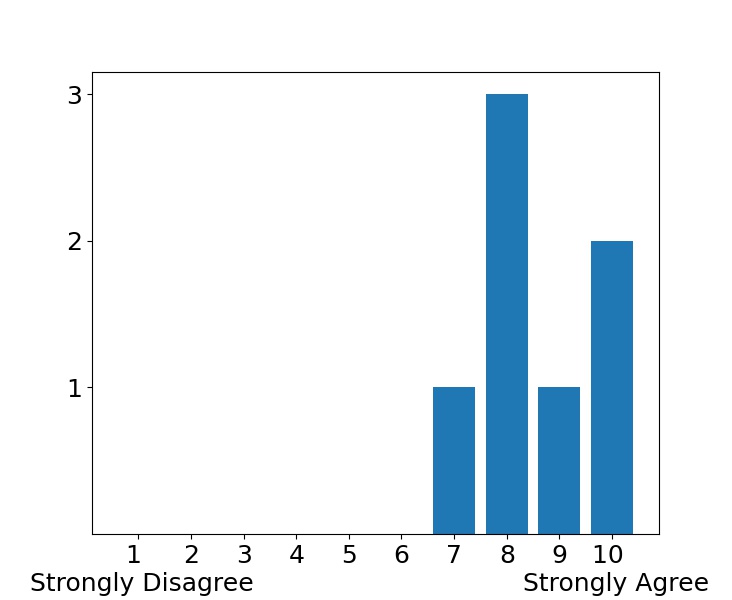}
\caption{Results from Question 15: "Scientific research is the type of research that is being done in faculty member research labs. Please indicate the extent you agree with the following statement: I conducted scientific research in the SES 294 course." On a scale from 1 to 10 with 1 being "strongly disagree" and 10 being "strongly agree," all students believed they had conducted scientific research to some extent in the course.} \label{fig:Q12:conductresearch}
\end{figure}

Question 15 requested students to rate on a scale from 0 to 10, with 1 being strongly disagree, and 10 being strongly agree, the extent at which they conducted scientific research in the course (see Figure \ref{fig:Q12:conductresearch}). All students agreed that they "had conducted scientific research in the SES 294 course" with a score of at least 7. Students in general found that the research was important and useful and that the experience was indicative of how a professional astronomer conducts research. Question 16 asked students to elaborate in paragraph form their response to Question 15. One response stated, \textit{"I was able to work in a team the entire time and be apart of a group. Every step of research from planning to actual observing we as a group got to participate in and actually get a real taste of astronomical research in a real setting. In encourages me to keep the path and continue to one day be apart of a research group myself."} Another response stated, \textit{"I feel the course was a mix of simulated research with some aspects of the course being in line with the instructors current research. Although I would have liked to be more of a part of the instructors research I believe in the short time we had everything I did and was taught is super important to my path of becoming a researcher. I've already used skills learned in [SES 294] to solve problems in [SES 394]."} There were no comments that were contradictory to the ratings provided in Question 15, and all responses indicated that research was performed in the course at varying degrees.

\subsection{Career Aspirations}

\begin{figure}[bt!] 
\centering
\includegraphics[width=\linewidth]{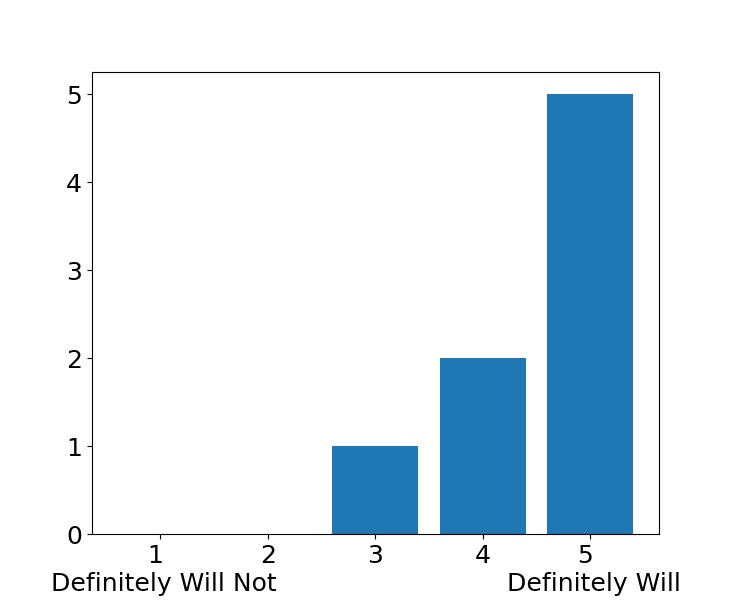}
\caption{Results from Question 17: "To what extent do you intend to pursue a science-related research career?" On a scale from 1 to 5, 87.5\% of students (extent of 4 or higher) intend to pursue a scientific career in the future.}\label{fig:Q16:pursuecareer}
\end{figure}

\begin{figure*}
\centering
\includegraphics[width=\textwidth]{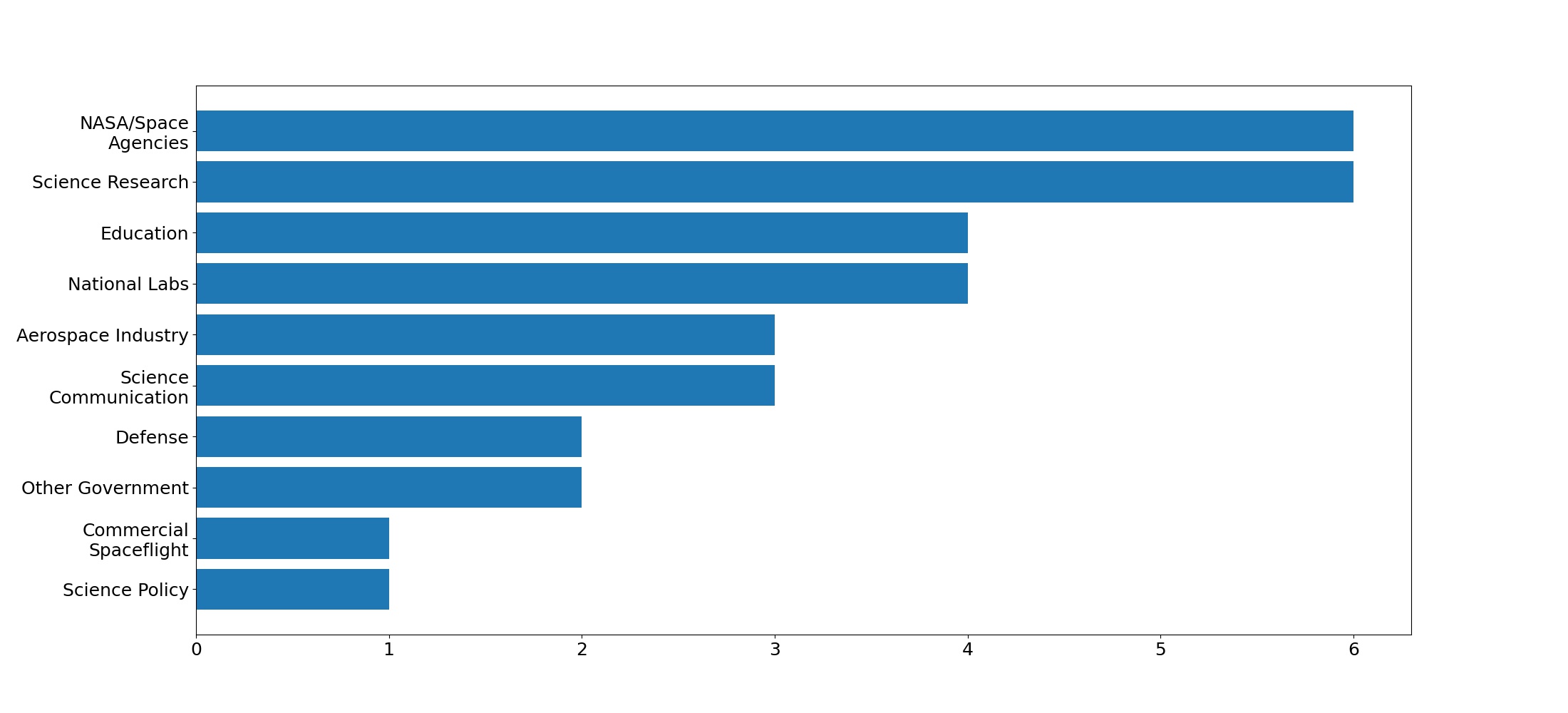}
\caption{Results from Question 18: "Which sectors most closely match your career goals?" with multiple options possible.
}\label{fig:Q17:careers}
\end{figure*}

Seven of eight students are likely to or definitely will pursue a science-related career (Questions 17 and 18, see Figure \ref{fig:Q16:pursuecareer}), and were highly interested in careers associated with science research, NASA/international space agencies, education, national laboratories, the aerospace industry, and science communication (see Figure \ref{fig:Q17:careers}). 



\subsection{Sense of Belonging}
\begin{figure}[bt!]
\centering
\includegraphics[width=0.7\linewidth]{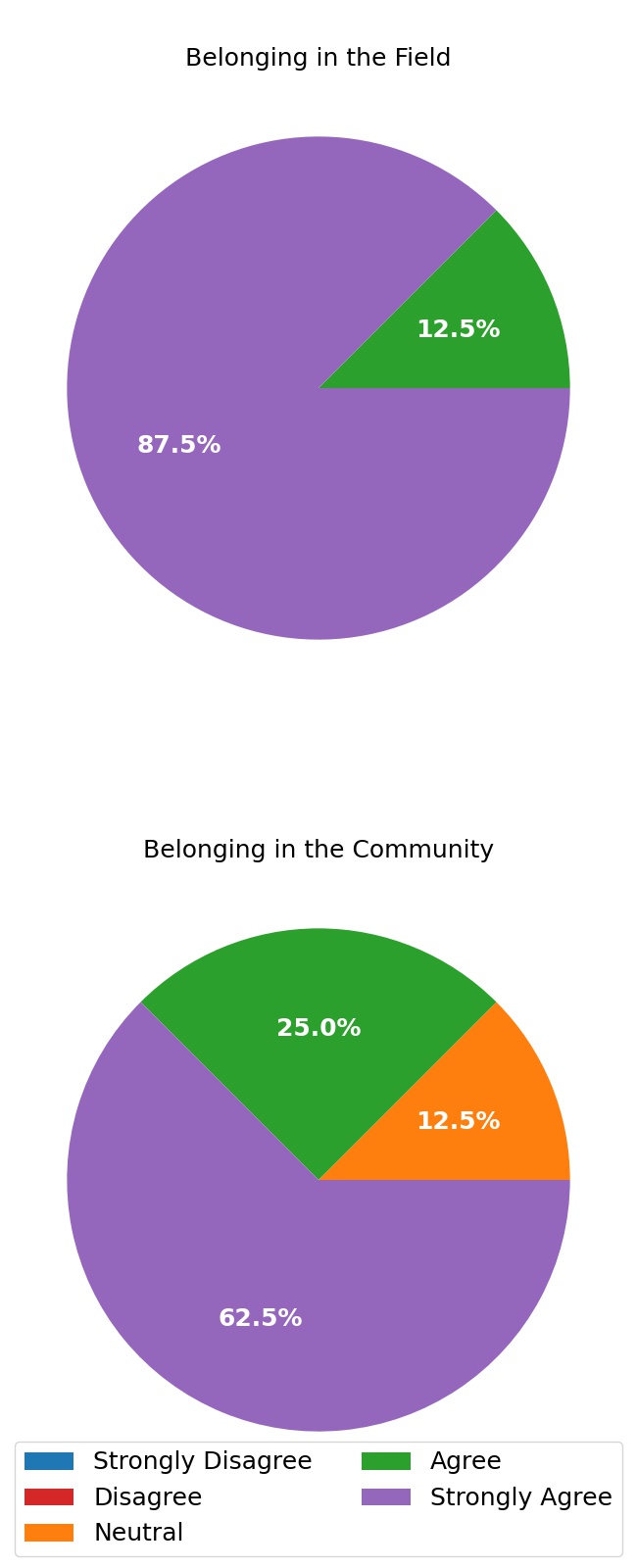}
\caption{Results from Question 19: "Please indicate the extent to which you agree with the statements below." The sub-questions were, "I feel like I belong in the field of astronomy" and "I have a strong sense of belonging to the community of astronomers." Sense of belonging is another important factor in overall student retention within the field of astronomy, and the responses collected from this survey suggest that students feel welcome witin the field and community.
}\label{fig:Q7:belonging}
\end{figure}

Two sub-questions to Question 19, shown in Figure \ref{fig:Q7:belonging}, asked students if they felt they belonged in the community of astronomers and the field of astronomy itself. Seven of eight students strongly agreed that they felt they belonged in the field of astronomy, and five of eight students strongly agreed that they belonged to the community of astronomers. 

\subsection{Demographic Information}

\begin{figure}[bt!]
\centering
\includegraphics[width=\linewidth]{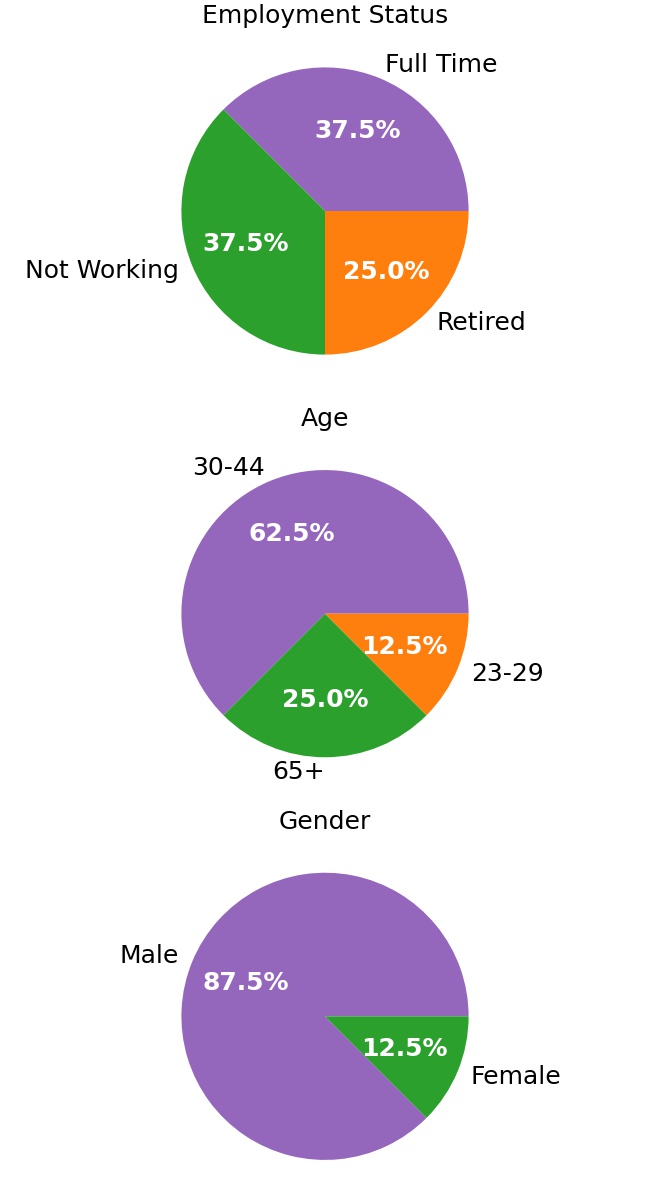}
\caption{Results from Questions 20, 22, and 23 which asked, "Are you currently working while completing your degree," age, and gender. Collected demographics as a whole suggest that this CURE enrolled students from non-traditional student backgrounds.
}\label{fig:Q18:demographics}
\end{figure}

Figure \ref{fig:Q18:demographics} shows the results for Questions 20, 22, and 23, all related to demographics. Students were roughly evenly split in identifying as full-time students, full-time workers, and retired. Of the full-time working students, they were employed in careers related to business, farming, and defense (Question 21). 62.5\% of students were in the 30-44 age group, consistent with the average age within the online APS program. Almost all students identified as male while only one respondent identified as female. While the distribution of gender varies significantly between our course and the APS program, our sample size is small comparatively. Addtionally, we report that the overall demographics presented in these survey results are representative of the demographics of the course as a whole, and are not skewed towards any particular demographic surveyed in these questions. Finally, although our sample size is small, with only single responses for some demographics question options, responses for questions cannot be linked between participants and all results are presented in aggregate.

\section{Interpretation} \label{sec: summary}
Students conveyed positive reception to many aspects of the course. In general, participating students found the pace and amount of content to be manageable, although some described the pace of the Python lessons specifically as too fast. Students generally found themselves discussing elements of the research project and course concepts with each other and the instructors often. The frequency of these performed actions in the course is encouraging, as these actions are common in a STEM research environments and contribute directly to the development of scientific skills and collaborative research \citep{lee2005impact}. Multiple respondents expressed a desire for a longer course length, which would allow for more time to learn and practice Python coding, have more evenly distributed course content, and to go more into depth with certain key concepts.

The responses to the course impact survey questions were encouraging, suggesting that there is evidence that the course positively impacted the students' abilities to participate in opportunities related to astronomy research. In particular, collaborative work and faculty interaction are critical components to pursuing a research-related career.
Students appear highly interested in pursuing astronomy and space science-related careers, and felt as though they were acquiring the skills and experiences necessary to succeed in those careers.
Additionally, the overall impression of useful skills learned in the CURE demonstrates that the students had an understanding of what tools are useful in future courses and research programs.

Students reported exceptional confidence in multiple skills taught in the course, especially pertaining to data analysis with spreadsheet software and online/GUI tools for observation preparation and image analysis. Collaborative Google spreadsheets were the primary tool in which students performed calculations and created graphs related to the research projects, and was one of the first tools introduced in the class. 
Students were relatively less confident in writing science reports. 
Using science literature to guide research was used in a limited capacity in the class, serving to reinforce taught concepts and to assist in writing the project report. Due to the short nature of the class, there may not have been sufficient time to develop the process of writing scientific reports. Additionally, as there was only one report assigned which was due at the end of the course, students may not have readily seen feedback regarding their report submissions. Finally, reports were written as a group, and it is unclear to what extent contributions and understanding were evenly divided amongst group members. In future iterations of this course, possible avenues to address these comments include multiple phases of feedback on the reports throughout the course, changing the writing of reports to be individual rather than collaborative, and/or indicating authorship in certain sections.

Confidence in understanding data/observations could be attributed to observation run planning and the observation run attended over Zoom at the end of the course. 
Finally, a much wider spread in confidence was evident in various skills involving Python coding. 
The lack of confidence in navigating Python coding was likely due to the accelerated nature of the Python workshops at the end of the course. Only four class sessions (approximately 8 hours) out of 16 were dedicated to giving an introduction to the basics of Python coding, and all workshops were consecutive in the last four sessions (two weeks) of the course. There were no strict assignments dedicated to practicing Python, although the Python notebooks written were designed to give students sections in which to practice various concepts in Python coding. In future iterations of the course, a potential solution would be to introduce Python coding much earlier and to have more time dedicated in general to learning and practicing. Practice in spreadsheet calculations and plotting can be simultaneous with coding and can give students multiple options for which to perform their data analyses. One response in Question 7 in particular highlighted the need to more evenly distribute Python curriculum throughout the semester, stating \textit{"I would have preferred to learn the Python coding a little by little each day, including regular homework throughout the semester since that is of particular interest to anyone in the field."} Regardless, at least one response indicated the importance of Python coding in preparing them for future courses, stating that \textit{"coding is completely new to me, but this leaves me better prepared for other classes I must take."}

All respondents felt a great sense of belonging to the field and community of astronomers, that the daily life of an astronomer was appealing, and that working on a team conducting research was personally satisfying. 
Relatively fewer students strongly felt that they identified as an astronomer, although no students disagreed with any of statements regarding science identity. A majority of students strongly identified with community values associated with scientists and astronomers, believing that "discussing new theories and ideas between scientists is important," "discovering something new in the sciences is thrilling," and that it is "valuable to conduct research that builds on the world's scientific knowledge." 
These results are encouraging, as they suggest that the students have a strong desire to continue pursuing astronomy research. 
Students mostly agreed that they conducted scientific research in the course, and that the course was indicative of how professional astronomers conduct research. 

Students were also very motivated to pursue scientific careers, with particular interest towards science research, space agencies, education, the aerospace industry, and science communication. These results are encouraging, as they suggest that the students not only gained experience in the tools/skills necessary for research programs, but that they also garnered an enthusiasm and motivation to pursue and conduct research in the future. Additionally, students recognized the importance of collaboration in research, and were able to practice collaborative interaction with peers and instructors.

The results of the survey are particularly important for retention within the online APS program and future ventures in astronomy research programs. Sense of belonging in particular has been found to be important in persistence towards degree attainment and is related to positive interactions with peers and faculty \citep{astin1993}. The positive attitude towards belonging based on the responses to these questions supports the notion that most of these students would be willing to pursue further studies and career opportunities in astronomy.

Responses to demographic questions revealed a mostly equal distribution of full-time employed, full-time student, and retired individuals. Students working full-time were employed in a variety of sectors, including business, farming, and defense. A majority of students were male and/or within the 30-44 age group. The results of this limited set of demographic information highlights an aspect of online education that is unique: equitable access to astronomy education for individuals that do not fit in with the typical in-person bachelor degree-seeking student mold.

\section{Conclusions and Future Work} \label{sec: conclusion}
We have developed a rigorous online CURE on observational astronomy for online degree-seeking astronomy students at ASU. The goal of this CURE was to expose students to the practice of conducting astronomical research by taking part in a research project being performed by Authors J. Hom and J. Patience. Students learned about telescopes, astronomical instruments, adaptive optics systems, observing run planning, and image viewing/processing tools. Students worked together in groups to analyze astronomical images of A-stars taken with the NIRC2 instrument at Keck Observatory, identifying candidate binary companions and measuring their positions and brightnesses. Students utilized the online and collaborative spreadsheet software Google Sheets to record their measurements and conduct their analyses, with a later introduction to Python coding as a way to perform similar tasks. The course concluded with students attending an observing run via Zoom held at Keck Observatory, where author J. Patience, collaborators, and observatory staff conducted additional observations for the research project.

A post-course survey distributed to students reported the CURE was well-received across multiple aspects including skills development and personal impact. The most common feedback received about the course suggested course content be more evenly distributed, and that the course itself should be scheduled as a longer course. Approval has already been granted to extend the length of the course from six to eight weeks in future iterations. A second look at the curricular structure and organization will be performed so that course concepts may be distributed more evenly. Students also requested the Python curriculum to be more evenly distributed throughout the course. In future iterations of the course, a possible avenue to pursue would be the simultaneous teaching and use of spreadsheet software and Python coding, allowing students to understand the advantages and disadvantages of using either in data analyses. This would also allow a longer period of time over which students could practice writing scripts and functions in Python with instructor guidance and assistance.

Due to the nature of this study, we were unable to perform comparisons of student responses to these questions at different stages of the course, and that the results of our survey may be somewhat ambiguous as a result. In future studies of this CURE and similar courses, surveys will be distributed at the beginning and end of the course in order to assess the evolution of student perspectives, confidences, and experiences as the course progresses. 
Regardless, the mostly positive quantitative reception along with the very positive qualitative feedback still suggest that the CURE made a positive impact on students' engagement and interest in the field of astronomy, and participants were enthusiastic about continuing their studies in astronomy and seeking out science-related fields, consistent with what was found in \cite{Hewitt2023}, which contained a pre- and post-survey along with a small survey response sample. 
Coming from various age and employment backgrounds, this course and others like it present a unique opportunity in which students have equitable access to astronomy research and the tools necessary to pursue astronomy and other STEM related careers in the future.

\section{Availability of supporting data and materials}

In accordance to the IRB-approved survey protocol, the survey responses linked to individual anonymized participants are not available, due to the small sample size and potential to identify individuals. Consequently, only the aggregate data per survey question are provided in the figures throughout the paper and in tabular form in Appendix \ref{appendix:fullsurveyresults}.







\subsection{List of abbreviations}
\begin{itemize}
    \setlength\itemsep{0.5em}
    \item Alt/Az -- Altitude/Azimuth
    \item AO -- Adaptive Optics
    \item ASU -- Arizona State University
    \item H-R Diagram -- Hertzsprung-Russell Diagram
    \item IRB -- Institutional Review Board
    \item NASA ADS -- National Aeronautics and Space Administration Astrophysical Data System, \url{https://ui.adsabs.harvard.edu}
    \item OURS -- Online Undergraduate Research Scholars, \url{ https://ours.thecollege.asu.edu}
    \item RA/Dec -- Right Ascension/Declination
    \item SAOimageDS9 -- An image display and visualization tool for astronomical data developed by Smithsonian Astrophysical Observatory, \url{ https://sites.google.com/cfa.harvard.edu/saoimageds9?pli=1}
    \item SES 294 -- Course number for class {\it Research Experience in Astronomical Imaging} offered in the School of Earth and Space Exploration at Arizona State University as an elective in the online degree program {\it Astronomical and Planetary Sciences}
    \item SIMBAD -- SIMBAD astronomical database, \cite{wenger2000};  \url{http://simbad.cds.unistra.fr/simbad}
\end{itemize}

\subsection{Ethical Approval}
Prior to submission of the Institutional Review Board survey protocol application, all authors completed the {\it Human Research}, {\it IRB - Social \& Behavioral Research}, and {\it Basic Course} training courses administered online through the Collaborative Institutional Training Initiative. The survey protocol was approved by the Institutional Review Board of the Office of Research Integrity and Assurance (ORIA) at ASU (Approval Protocol Title: Survey of SES 294 Students from Summer 2022 B Session). The survey conditions included informing the students that participation in the survey was optional and individual questions could be skipped, obtaining consent, and not releasing individualized responses due to the limited number of students in the class. The conditions in the IRB were followed in the survey and study.





\subsection{Consent for publication}

As part of the IRB-approved survey, the students who participated in the study confirmed consent for the data to be published. Only aggregate data are reported in the figures and tables and made available in Appendix \ref{appendix:fullsurveyresults} in order to protect the anonymity of the individuals who responded to the survey.



\subsection{Competing Interests}

The authors declare that they have no competing interests. 

\subsection{Funding}

University seed funding from the OURS Program was provided for the SES 294 class. As an internal award, there is no award/grant number. The Co-Principal Investigators of the award were co-authors J. Patience and K. Knierman. The funding supported the teaching assistant (lead author J. Hom) and enabled the travel of a research student to the Keck observing run to serve as the liaison to the SES 294 students on the Zoom session. 

\subsection{Author's Contributions}
\begin{itemize}
    \item \textbf{Justin Hom:} conceptualization, methodology, software, formal analysis, investigation, resources, data curation, writing--original draft, visualization, project administration
    \item \textbf{Jennifer Patience:} conceptualization, methodology, formal analysis, investigation, resources, writing--original draft, project administration, supervision, funding acquisition
    \item \textbf{Karen Knierman:} conceptualization, methodology, investigation, resources, writing--review \& editing, project administration, supervision
    \item \textbf{Molly Simon:} writing--review \& editing
    \item \textbf{Ara Austin:} writing--review \& editing
\end{itemize}


\section{Acknowledgements}

The authors would like to acknowledge the students of the SES 294 course, along with Professor Maitrayee Bose for providing a guest presentation on her research and openings for research positions in her group. We would also like to acknowledge M.S. Jasmine Garani, graduate students Marah Brinjikji, Adam Smith, and Emma Softich, and Professor Eric Nielsen for their involvement in the research project and observing run at Keck Observatory. Furthermore, we would like to thank and acknowledge Keck Observatory staff for answering questions from our students during the observing run.

\bibliography{paper-refs}

\appendix
\section{Appendix: Full Survey Results} \label{appendix:fullsurveyresults}
Here we report the full survey results in Tables \ref{tab:Q1data}-\ref{tab:Qbelongingdata}. Questions 6-8 and 16 are not reported on as they are long form paragraph responses. Questions 20-23 are not reported on due to the small sample size ($N=8$) and is potentially identifying/non-anonymous.


\begin{table*}[b]
\caption{1. How would you describe your interest in topics/skills that you learned in the course?}\label{tab:Q1data}
\begin{tabularx}{\linewidth}{c c c c c}
\toprule
 {High Disinterest (1)} & {Moderate Disinterest} & {Average Interest} & {Moderate Interest} & {High Interest (5)}\\
\midrule
0 & 0 & 0 & 0 & 8  \\


\bottomrule
\end{tabularx}

\begin{tablenotes}
\item Eight total responses.
\end{tablenotes}
\end{table*}

\begin{table*}[bt!]
\caption{2. How would you describe the pace of the class overall?}\label{tab:Q2data}
\begin{tabularx}{\linewidth}{c c c c c}
\toprule
 {Too Slow (1)} & {A Little Slow} & {Manageable} & {A Little Fast} & {Too Fast (5)}\\
\midrule
0 & 0 & 8 & 0 & 0  \\


\bottomrule
\end{tabularx}

\begin{tablenotes}
\item Eight total responses.
\end{tablenotes}
\end{table*}

\begin{table*}[bt!]
\caption{3. How would you describe the pace of the Python notebook workshops?}\label{tab:Q3data}
\begin{tabularx}{\linewidth}{c c c c c}
\toprule
 {Too Slow (1)} & {A Little Slow} & {Manageable} & {A Little Fast} & {Too Fast (5)}\\
\midrule
0 & 0 & 4 & 2 & 2  \\


\bottomrule
\end{tabularx}

\begin{tablenotes}
\item Eight total responses.
\end{tablenotes}
\end{table*}

\begin{table*}[bt!]
\caption{4. How would you describe the amount of class material?}\label{tab:Q4data}
\begin{tabularx}{\linewidth}{c c c c c}
\toprule
 {Too Much (1)} & {Somewhat Much} & {Manageable} & {Somewhat Little} & {Too Little (5)}\\
\midrule
0 & 0 & 7 & 1 & 0  \\


\bottomrule
\end{tabularx}

\begin{tablenotes}
\item Eight total responses.
\end{tablenotes}
\end{table*}

\begin{table*}[bt!]
\caption{5. How often did you do the following in SES 294? }\label{tab:Q11data}
\begin{tabularx}{\linewidth}{L c c c c}
\toprule
 {Subquestion} & {Never (1)} & {Once or Twice} & {Once or Twice a Week} & {Multiple Times a Week (4)}\\
\midrule
discuss elements of my investigation with my classmates or instructors & 0 & 0 & 1 & 7  \\
reflect on what I was learning & 0 & 0 & 1 & 7 \\
contribute my ideas and suggestions during class discussions & 0 & 0 & 3 & 5 \\
provide constructive criticism to classmates and challenge each other’s interpretations & 0 & 1 & 2 & 5 \\
share the problems I encountered during my investigation and seek input on how to address them & 0 & 0 & 1 & 7 \\


\bottomrule
\end{tabularx}

\begin{tablenotes}
\item Eight total responses.
\end{tablenotes}
\end{table*}

\begin{table*}[bt!]
\caption{9. How much do you think this SES 294 class impacted you in these aspects? }\label{tab:Q8data}
\begin{tabularx}{\linewidth}{L c c c c c}
\toprule
 {Subquestion} & {Poor (1)} & {Slightly Poor} & {Average} & {Good} & {Excellent (5)}\\
\midrule
Skills development & 0 & 0 & 0 & 3 & 5  \\
Ability to work in a research group & 0 & 0 & 0 & 0 & 8  \\
Confidence to approach faculty with questions & 0 & 0 & 0 & 0 & 8  \\
Motivation to work in a related field & 0 & 0 & 0 & 1 & 7  \\


\bottomrule
\end{tabularx}

\begin{tablenotes}
\item Eight total responses.
\end{tablenotes}
\end{table*}

\begin{table*}[bt!]
\caption{10. What skills from this class do you see yourself using again in the future? (Check all that apply.)}\label{tab:Q9data}
\begin{tabularx}{\linewidth}{L c}
\toprule
 {Option} & {Count}\\
\midrule
Using spreadsheets for calculations (e.g., Excel or Google Sheets) & 7 \\
Using ds9 for examining astronomical images & 8 \\
Using Python or other coding & 8 \\
Using Slack or other messaging services to facilitate teamwork & 8 \\
Using Zoom for workshop style learning & 7 \\
Other* & 2 \\


\bottomrule
\end{tabularx}

\begin{tablenotes}
\item Eight total responses. *Other responses: \textit{"Reading multiple scientific publications to look for relevant information to my research. Searching archived celestial objects observation data. Planning of future observations (target locations in sky on planned observation nights and affects of Moon location)," "simbad, staralt and nasa archives"}.
\end{tablenotes}
\end{table*}

\begin{table*}[bt!]
\caption{11. Please indicate how confident you are in your ability with the following:}\label{tab:Q5data}
\begin{tabularx}{\linewidth}{L c c c c c}
\toprule
 {Subquestion} & {Not Confident at All (1)} & {Slightly Confident} & {Moderately Confident} & {Very Confident} & {Absolutely Confident (5)}\\
\midrule
to use technical science skills (use of tools, instruments, and/or techniques) & 0 & 0 & 0 & 3 & 5  \\
to figure out what data/observations to collect and how to collect them & 0 & 0 & 0 & 3 & 5  \\
to use scientific literature and/or reports to guide research & 0 & 0 & 0 & 3 & 5  \\
Google Sheets/Plotting & 0 & 0 & 0 & 1 & 7  \\
Tools such as: Staralt/SIMBAD/ds9 & 0 & 0 & 0 & 2 & 6  \\
ASU Library \& reference searches & 0 & 0 & 0 & 3 & 5  \\
Write a Project Report & 0 & 0 & 4 & 2 & 2  \\


\bottomrule
\end{tabularx}

\begin{tablenotes}
\item Eight total responses.
\end{tablenotes}
\end{table*}

\begin{table*}[bt!]
\caption{12. Please indicate how confident you are in your ability to do the following using the Python programming language:}\label{tab:Q6data}
\begin{tabularx}{\linewidth}{L c c c c c}
\toprule
 {Subquestion} & {Not Confident at All (1)} & {Slightly Confident} & {Moderately Confident} & {Very Confident} & {Absolutely Confident (5)}\\ 
\midrule
Use Python Notebooks & 1 & 0 & 5 & 1 & 1  \\
assign variables a value & 1 & 1 & 1 & 3 & 2  \\
arithmetic with variables & 1 & 1 & 1 & 4 & 1  \\
create and manipulate arrays & 1 & 3 & 2 & 1 & 1  \\
create plots & 1 & 2 & 0 & 3 & 2  \\


\bottomrule
\end{tabularx}

\begin{tablenotes}
\item Eight total responses.
\end{tablenotes}
\end{table*}

\begin{table*}[bt!]
\caption{13. Please indicate the extent to which you agree with the statements below }\label{tab:Q7data}
\begin{tabularx}{\linewidth}{L c c c c c}
\toprule
 {Subquestion} & {Strongly Disagree (1)} & {Disagree} & {Neutral} & {Agree} & {Strongly Agree (5)}\\
\midrule
I derive great personal satisfaction from working on a team that is doing important research & 0 & 0 & 0 & 0 & 8  \\
I have come to think of myself as an astronomer & 0 & 0 & 1 & 4 & 3  \\
The daily work of an astronomer is appealing to me & 0 & 0 & 0 & 0 & 8  \\


\bottomrule
\end{tabularx}

\begin{tablenotes}
\item Eight total responses.
\end{tablenotes}
\end{table*}

\begin{table*}[bt!]
\caption{14. Please rate how much the person in the description is like you.
 }\label{tab:Q10data}
\begin{tabularx}{\linewidth}{L c c c c c c}
\toprule
 {Subquestion} & {Not like me at all (1)} & {Not like me} & {A little like me} & {Somewhat like me} & {Like me} & {Very much like me (6)}\\
\midrule
A person who thinks it is valuable to conduct research that builds on the world’s scientific knowledge & 0 & 0 & 0 & 0 & 1 & 7  \\
A person who feels discovering something new in the sciences is thrilling & 0 & 0 & 0 & 0 & 1 & 7  \\
A person who thinks discussing new theories and ideas between scientists is important & 0 & 0 & 0 & 0 & 0 & 8  \\


\bottomrule
\end{tabularx}

\begin{tablenotes}
\item Eight total responses.
\end{tablenotes}
\end{table*}

\begin{table*}[bt!]
\caption{15. Scientific research is the type of research that is being done in faculty member research labs. Please indicate the extent you agree with the following statement: I conducted scientific research in the SES 294 course.
 }\label{tab:Q12data}
\begin{tabularx}{\linewidth}{c c c c c c c c c c}
\toprule
Strongly Disagree & & & & & & & & & Strongly Agree\\
{1} & {2} & {3} & {4} & {5} & {6} & {7} & {8} & {9} & {10}\\
\midrule
0 & 0 & 0 & 0 & 0 & 0 & 1 & 3 & 1 & 2  \\


\bottomrule
\end{tabularx}

\begin{tablenotes}
\item Seven total responses.
\end{tablenotes}
\end{table*}

\begin{table*}[bt!]
\caption{17. To what extent do you intend to pursue a science-related research career?
 }\label{tab:Q17data}
\begin{tabularx}{\linewidth}{c c c c c}
\toprule
Definitely Will Not & & & & Definitely Will\\
{1} & {2} & {3} & {4} & {5} \\
\midrule
0 & 0 & 1 & 2 & 5  \\


\bottomrule
\end{tabularx}

\begin{tablenotes}
\item Eight total responses.
\end{tablenotes}
\end{table*}

\begin{table*}[bt!]
\caption{18. Which sectors most closely match your career goals? (more than one answer is possible))}\label{tab:Q18data}
\begin{tabularx}{\linewidth}{L c}
\toprule
 {Option} & {Count}\\
\midrule
Education & 4 \\
NASA/Other International Space Agencies & 6 \\
National Labs & 4 \\
Other Government & 2 \\
Science Research & 6 \\
Aerospace Industry & 3 \\
Commercial Spaceflight & 1 \\
Defense & 2 \\
Science Communication & 3 \\
Data Science & 0 \\
Science Policy & 1 \\
Medical & 0 \\
Tech Industry & 0 \\
Retail & 0 \\
Business & 0 \\
Other commercial/industry & 0 \\
Other* & 3 \\


\bottomrule
\end{tabularx}

\begin{tablenotes}
\item Eight total responses. *Other responses: \textit{"My top 3 goals are to do frontline astronomy research, give back to the next generation of astronomers and educate the general public," "As a retiree, I was looking at this course as an opportunity to understand the processes that are active in science, today. My only goal was to understand and educate myself, not to go into a career," "writing about science and astronomy for magazines and documdentaries."}.
\end{tablenotes}
\end{table*}

\begin{table*}[bt!]
\caption{19. Please indicate the extent to which you agree with the statements below: }\label{tab:Qbelongingdata}
\begin{tabularx}{\linewidth}{L c c c c c}
\toprule
 {Subquestion} & {Strongly Disagree (1)} & {Disagree} & {Neutral} & {Agree} & {Strongly Agree (5)}\\
\midrule
I have a strong sense of belonging to the community of astronomers & 0 & 0 & 1 & 2 & 5  \\
I feel like I belong in the field of astronomy & 0 & 0 & 0 & 1 & 7  \\


\bottomrule
\end{tabularx}

\begin{tablenotes}
\item Eight total responses.
\end{tablenotes}
\end{table*}

\end{document}